\documentclass[aps,pre,epsf,twocolumn,floatfix]{revtex4}
\usepackage{amsmath}
\usepackage{amssymb}
\usepackage{epsfig}
\usepackage{graphicx}
\usepackage[T1]{fontenc}
\usepackage[utf8]{inputenc}
\usepackage{color}
\usepackage[normalem]{ulem}
\usepackage{tikz}
\usetikzlibrary{positioning}

\usepackage{comment}

\graphicspath {{./figures/}}
\makeatletter
\def\input@path{{./figures/}}
\makeatother

\begin{document}

\title{Two-dimensional dilute Baxter-Wu model: Transition order and universality}

\author{A.~R.~S. Mac\^edo$^{1,2}$}

\author{A. Vasilopoulos$^3$}

\author{M. Akritidis$^3$}

\author{J.~A. Plascak$^{1,4,5}$}

\author{N.~G. Fytas$^{3}$}
\email{nikolaos.fytas@coventry.ac.uk}

\author{M. Weigel$^6$}

\affiliation{$^1$Departamento de F\'\i sica, Instituto de Ci\^encias Exatas, Universidade Federal de Minas 
Gerais, C.P. 702. 30123-970 Belo Horizonte, MG - Brazil}

\affiliation{$^2$Instituto Federal do Maranh\~ao - Campus Imperatriz,  65919-050, Imperatriz, MA - Brazil}

\affiliation{$^3$Centre for Fluid and Complex Systems, Coventry
	University, Coventry CV1 5FB, United Kingdom}

\affiliation{$^4$Universidade Federal da Para\'iba, Centro de Ci\^encias Exatas e da Natureza - Campus I, 
Departamento de F\'isica - CCEN Cidade Universit\'aria 58051-970 Jo\~ao Pessoa, PB - Brazil}

\affiliation{$^5$Center for Simulational Physics, University of Georgia, 30602 Athens, GA - USA }


\affiliation{$^6$Institut für Physik, Technische Universität Chemnitz, 09107 Chemnitz, Germany}

\date{\today}

\begin{abstract}
We investigate the critical behavior of the two-dimensional spin-$1$ Baxter-Wu model in the presence of a crystal-field coupling $\Delta$ with the goal of determining the universality class of transitions along the second-order part of the transition line as one approaches the putative location of the multicritical point. We employ extensive Monte Carlo simulations using two different methodologies: (i) a study of the zeros of the energy probability distribution, closely related to the Fisher zeros of the partition function, and (ii) the well-established multicanonical approach employed to study the probability distribution of the crystal-field energy. A detailed finite-size scaling analysis in the regime of second-order phase transitions in the $(\Delta, T)$ phase diagram supports previous claims that the transition belongs to the universality class of the $4$-state Potts model.
For positive values of $\Delta$, we observe the presence of strong finite-size effects, indicative of crossover effects due to the proximity of the first-order part of the transition line. Finally, we demonstrate how a combination of cluster and heat-bath updates allows one to equilibrate larger systems, and we demonstrate the potential of this approach for resolving the ambiguities observed in the regime of $\Delta \gtrsim 0$.  
\end{abstract}

\maketitle

\section{Introduction}
\label{sec:Intro}

Most of the commonly studied spin models of statistical mechanics such as the Ising and Potts or O($n$) models are spin-inversion symmetric. A notable exception to this rule is the Baxter-Wu (BW) model~\cite{baxter73,baxter_book} that was originally introduced by Wood and Griffiths 
\cite{wood72,merlini72}. The commonly studied version is defined on the triangular lattice with $N$ sites and the Hamiltonian function
\begin{equation}
{\cal{H}}_{\rm BW}=-J\sum_{\langle ijk\rangle}{\sigma_i}{\sigma_j}\sigma_k, 
\label{eq:Ham}
\end{equation}
where $J > 0$ denotes a ferromagnetic exchange coupling. The sum $\langle ijk\rangle$ extends over all elementary triangles,  and 
$\sigma_i=\pm1$ are Ising like spin-$1/2$ variables. The presence of three-spin interactions leads to the mentioned violation of spin-inversion symmetry, and it results in a four-fold degeneracy of the ground state:  there is one ferromagnetic state with all spins up, and three ferrimagnetic states with down spins in two sublattices and up spins in the third.The triangular lattice can be decomposed into three sublattices, A, B, and C, as shown in Fig.~\ref{fig:lattice} below. Note that the model of Eq.~(\ref{eq:Ham}) is self-dual~\cite{wood72,merlini72}, resulting in the same critical temperature
as that of the spin-$1/2$ Ising model on the square lattice, i.e.,
$k_{\rm B}T_{\rm c}/J =2/\ln{(\sqrt{2}+1)} = 2.269185\cdots$, where $k_{\rm B}$ denotes Boltzmann's constant. 

\begin{figure}
\includegraphics[width=0.84\hsize]{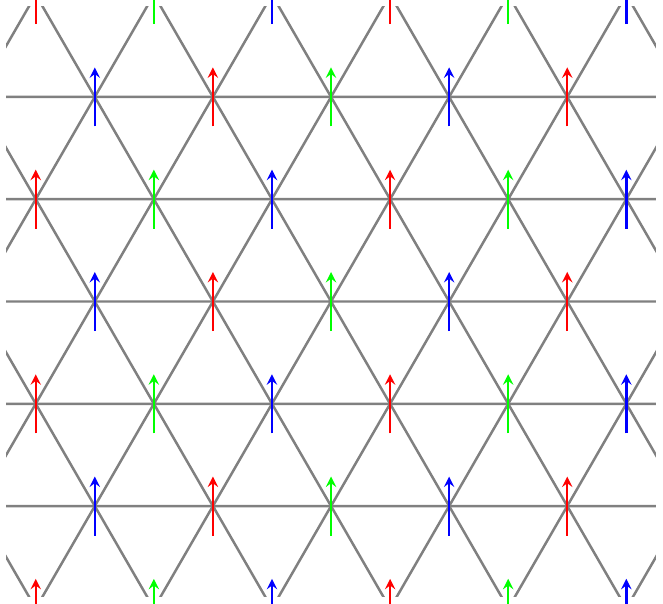}
	\caption{Representation of the triangular lattice of the Baxter-Wu model as a superposition of three sublattices, A, B, and C. Each sublattice corresponds to spins of different color. The spins are shown in the ferromagnetic ground state.}  
		\label{fig:lattice}
\end{figure}

An exact solution of the model was provided early on by Baxter and Wu~\cite{baxter73,baxter_book}, supplying the critical exponents $\alpha = 2/3$, $\nu = 2/3$, and $\gamma = 7/6$. In the following, it was also shown that its critical behavior corresponds to a conformal field theory with central charge $c = 1$~\cite{alcaraz97,alcaraz99}. Due to the four-fold symmetry of the ground state, it is expected that the critical behavior of the $q = 4$ model Potts belongs to the universality class of the Baxter-Wu model~\cite{domany78}. While, therefore, the critical exponents of the two models are identical, the same does not apply to the scaling corrections: the $4$-state Potts model exhibits logarithmic corrections with the system size~\cite{wu82}, whereas the Baxter-Wu model has power-law corrections with a correction-to-scaling exponent $\omega = 2$~\cite{alcaraz97,alcaraz99}. Recently, the model has attracted renewed attention, and various aspects of its critical behavior have been studied in substantial detail~\cite{hadjiagapiou05,shchur10,velonakis13,capponi14,velonakis18,jorge19,cavalcante19,liu22,monroe22}.

A natural generalization of the Baxter-Wu model~(\ref{eq:Ham}) 
results from the consideration of \emph{three} spin orientations $\sigma_{i} =  \{-1,0,1\}$ and the inclusion of an extra crystal-field (or single-ion anisotropy) coupling $\Delta$. The resulting Hamiltonian then reads
\begin{equation}\label{eq:Ham2}
  \mathcal{H}
  = -J\sum_{\langle ijk \rangle}\sigma_{i}\sigma_{j}\sigma_{k}+\Delta\sum_{i}\sigma_{i}^{2} = E_{J}+\Delta E_{\Delta},
\end{equation}
where $E_J$ and $E_\Delta$ denote the contributions of the exchange and the crystal field, respectively, to the total energy. Note that in the following we will use reduced units where $J=1$ as well as $k_{\rm B} = 1$. This choice of units follows the practise of some of the present authors to implement multicanonical simulations on spin-$1$ Blume-Capel and Baxter-Wu models, see the discussion in Sec.~\ref{sec:muca}. Although still rather simple, 
for this spin-$1$ model there exists no exact solution, except  for the case $\Delta \rightarrow-\infty$, where only configurations with $\sigma_{i} = \pm 1$ are allowed and the pure spin-$1/2$ Baxter-Wu model is recovered, and also at zero temperature, where the four ordered phases coexist with the paramagnetic phase in a multiphase point at $\Delta/J=2$ (accordingly, no transition is observed for $\Delta/J>2$). 

\begin{figure}[ht]
\includegraphics[width=0.99\hsize]{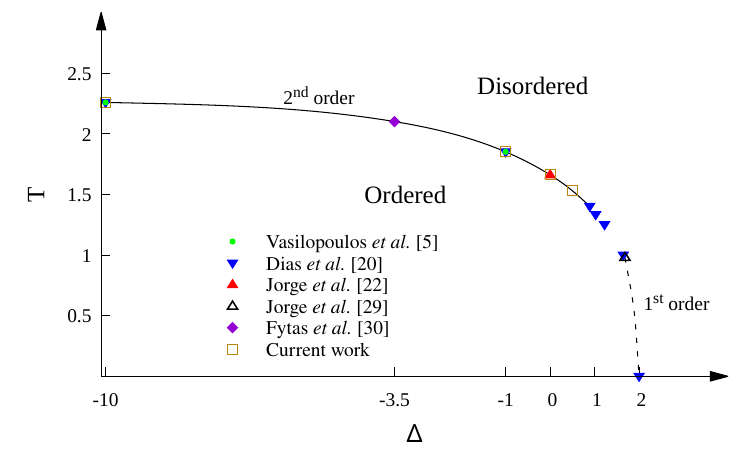}
\caption{Phase diagram of the two-dimensional spin-$1$ Baxter-Wu model including several estimates of transition points. The black dashed and continuous lines correspond to first- and second-order phase transitions, respectively. The intermediate regime between the two pentacritical point estimations by Dias \emph{et al.}~\cite{dias17} $(\Delta_{\rm pp}, T_{\rm pp}) \approx (0.8902, 1.4)$ and Jorge \emph{et al.}~\cite{jorge21} $(\Delta_{\rm pp}, T_{\rm pp}) \approx (1.68288(62), 0.98030(10))$ is not crossed by a line as its status is currently unclear.}
\label{fig:phase_diagram}
\end{figure}

Based on the analogy between the Baxter-Wu and the 
diluted Potts model~\cite{nienhuis79}, but also on a series of more recent results~\cite{dias17,costa04,jorge21}, it is now well established that the phase diagram of the spin-$1$ Baxter-Wu model in the $(\Delta, T)$-plane includes a multicritical point separating first- from second-order transition regimes. 
This is in contrast to an earlier prediction by
finite-size-scaling applied to transfer-matrix calculations, where a continuous transition only occurs in the limit 
$\Delta \rightarrow-\infty$~\cite{kinzel81}. In this respect, the model resembles the well-known Blume-Capel ferromagnet~\cite{blume66}, which exhibits a phase diagram with ordered ferromagnetic and disordered paramagnetic phases separated by a transition line with first- and second-order segments (the latter in the Ising universality class) connected by a tricritical point, whose location is known with high accuracy~\cite{malakis10,zierenberg15,kwak15,zierenberg17}. In contrast, there is no consensus on the precise location of the multicritical point for the spin-$1$ Baxter-Wu model --- see Fig.~\ref{fig:phase_diagram} but also Fig.~5 of Ref.~\cite{jorge21} for a summary regarding the phase diagram. Along the first-order transition line of Fig.~\ref{fig:phase_diagram} three ferrimagnetic phases and one ferromagnetic one coexist with the paramagnetic phase, forming a quintuple line that arrives at a pentacritical point where all five phases become identical. 

In addition to the question of the location of the multicritical point, the reign of universality along the second-order segment of the transition line has been put into question. While some earlier results based on the transfer matrix and conformal invariance suggested a continuous variation of critical exponents with the crystal field along the second-order transition line~\cite{costa04}, more recent studies reported a match of the observed critical behavior with that of the $4$-state Potts model~\cite{dias17}. Some authors have also suggested the scenario of a mixed-order transition with both first-order and second-order properties~\cite{jorge20}. 
Recently, however, some clear-cut evidence for a simple, continuous transition in the universality class of the $4$-state Potts model in the regime of  $\Delta < 0$ could be provided based on a highly optimized combination of Wang-Landau simulations that cross the transition at constant $\Delta$ and multicanonical simulations operating at constant temperature $T$ ~\cite{fytas22,vasilopoulos22}, such that questions remain only in the regime $\Delta \gtrsim 0$.

In the present work we study the spin-$1$ Baxter-Wu model at several values of the crystal-field coupling that also reach into the regime $\Delta > 0$. To this end, we employ two complementary Monte Carlo schemes, a recently proposed variant of studying Fisher's partition function zeros \cite{fisher65} dubbed \emph{energy 
probability distribution zeros}~\cite{costa17}, and the multicanonical approach applied to the crystal-field energy~\cite{berg92,zierenberg15,fytas22,vasilopoulos22}. While the latter method is well established in the literature, the former was to date shown to be robust and useful in only a few cases, including some simple spin systems and 
polymer chains~\cite{costa17,costa19,rodrigues21}. In this respect, the purpose of the present work is twofold: firstly, to explore the scope and limitations of the method of energy probability distribution zeros for the more complicated spin-$1$ Baxter-Wu model that lacks the up-down symmetry and, secondly, to investigate the criticality and universality of the model specifically for $\Delta \ge  0$, but still below the proposed location of the multicritical point.

The rest of the paper is organized as follows. In Sec.~\ref{sec:zeros} we outline the method based on the energy probability distribution zeros and show results for both the pure spin-$1/2$ and for the spin-$1$ Baxter-Wu model, the latter for several values of the crystal-field coupling in the range $-10 \le \Delta \le 0.5$. In Sec.~\ref{sec:muca} we complement the outcomes of Sec.~\ref{sec:zeros} via extensive multicanonical simulations at fixed values of the temperature in the regime where $\Delta \geq 0$. Finally, in Sec.~\ref{sec:conclusions} we summarize the main findings of the current work, comparing the implemented methodologies in the light of some additional preliminary results at $\Delta = 0$ obtained via an efficient numerical scheme that mixes cluster and heat-bath updates.

\section{Energy probability distribution zeros}
\label{sec:zeros}

\subsection{Description and finite-size scaling}
\label{sec:EPD_description}

As was recently discussed in Refs.~\cite{costa17,costa19,rodrigues21}, the study of zeroes in the energy probability distribution (EPD) allows for a straightforward determination of critical temperatures and the shift exponent $\theta = 1/\nu$ while avoiding the need of computing traditional thermodynamic quantities, such as the susceptibility or the specific heat. The method of EPD zeros is closely related to the Fisher zeros of the canonical partition function $\cal Z$~\cite{fisher65}, expressed as
\begin{equation}
{\cal Z}=\sum_Eg(E)e^{-\beta E}=e^{-\beta \epsilon_0}\sum_{n=1}^{\cal N}g(E_{n})e^{-\beta n\epsilon},
\label{eq:zn}
\end{equation}
where $E$ is the energy of the system, $g(E)$ is the number of states having energy $E$ (degeneracy), and $\beta=1/T$. In the last part of Eq.~(\ref{eq:zn}) we assume a discrete energy spectrum $E = E_n=\epsilon_0+n\epsilon$, where $\epsilon_0$ is the ground-state energy and $\epsilon$ denotes the level spacing,  $n=0,1,2,\ldots,\mathcal{N}$. Fisher noted that since \eqref{eq:zn} is a polynomial in $y=e^{-\beta \epsilon}$, it has $N$ complex zeros and since $g(E_n) \ge 0$ none of them are real. However, on approaching the thermodynamic limit ${\cal N}\rightarrow\infty$, some zeros might approach the real axis, thus leading to a non-analyticity at $y_c=e^{-\beta_{\rm c} \epsilon}$ corresponding to the phase transition at the inverse critical temperature $\beta_{\rm c}$. Since the partition function is not so straightforward to sample in a Monte Carlo simulation, we consider a somewhat different formulation. To this end, we multiply the right hand side of Eq.~(\ref{eq:zn}) by $1=e^{-\beta_0 \epsilon_0}e^{+\beta_0 \epsilon_0}$ to obtain
\begin{equation}
{\cal Z}_{\beta} = e^{-\triangle\beta \epsilon_0}\sum_{n=1}^{\cal N}h_{\beta_0}(n)x^n,
\label{eq:zb}
\end{equation}
where $\beta_0$ is the inverse of some reference temperature, $\triangle\beta=\beta-\beta_0$, $h_{\beta_0}(n)=g(E_{n}) e^{-\beta_0 E_{n}}$, and 
$x=e^{-\triangle\beta\epsilon}$. Note that $h_{\beta_0}(n)$ is the unnormalized 
canonical energy probability distribution at $\beta_0$, and it can be easily estimated from an energy histogram through Monte Carlo simulations. As is easily seen, when $\beta_0=\beta_c$ the dominant zero of the EPD is located at the fixed value $x_{\rm c} =(1,0)$ in the complex plane, thus simplifying the analysis.

For the finite lattice systems of linear size $L$ that are amenable to numerical simulation, one can systematically follow the behavior of the  most dominant zero $x_L^{\ast}$ that is approaching the real axis at $x_{\rm c} = (1,0)$ as $L$ is increased. In this way, it is possible to use finite-size scaling arguments to retrieve the critical temperature as well as the critical 
exponent $\nu$. Specifically, the algorithm proposed in Ref.~\cite{costa17} for this purpose is as follows. We first choose a starting guess $\beta_0^{j=0}$ of the inverse transition temperature and then iterate through the following steps:
\begin{enumerate}
\item[(1)] Simulate the system at $\beta = \beta_0^j$ and construct a histogram $h_{\beta_0^j}$.
\item[(2)] Find all the zeros of the polynomial with coefficients given by $h_{\beta_0^j}$,
\emph{i.e.}, 
\begin{equation}
\label{eq:hzeros}
\sum_{n=1}^{\cal N}h_{\beta_0^{j}}(n)x^n=0.
\end{equation}
\item[(3)] Find the dominant zero $(x^{j})^{\ast}$. Then:
\begin{itemize}
\item[(i)] if $(x^{j})^{\ast}$ is close enough to the point $(1, 0)$, $x_{L}^{\ast} = (x^{j})^{\ast}$ and stop;
\item[(ii)] else, make
\begin{equation}
\label{eq:iter}
\beta_0^{j+1} = -\epsilon^{-1}\ln{ [\operatorname{Re}(x^{j})^{\ast}]} + \beta_0^j
\end{equation}
and return to step (1). 
\end{itemize}
\end{enumerate}

After setting a convergence criterion, one ends up with estimates $x_{L}^{\ast}$ for the dominant zeros and hence with pseudocritical temperatures $T_{L}^{\ast}$, resp.\ $\beta_{L}^{\ast}$.  
Previous numerical results for several spin systems of Ising, Potts, and Heisenberg  type, as well as for homopolymeric models, indicate that~\cite{costa17,costa19,rodrigues21}: (i) the choice of the starting temperature $\beta_0$ is largely irrelevant for arriving at the dominant zero $x_{L}^{\ast}$; (ii) for $\beta \approx \beta_c$, only states with a non-vanishing probability are relevant to the transition, thus allowing to define a cutoff, 
$h_{\rm cut}$, affecting the left- and right-hand side  margins of the energy distribution. Discarding 
configurations where $h_{\beta_0^j}(n) < h_{\rm cut}$ substantially reduces the degree of the polynomial, especially with increasing system size; and (iii) to further simplify the polynomial, one can rescale the histogram by setting its maximum value to unity so that $\operatorname{max} h_{\beta_0^j} = 1$, since an overall rescaling of the partition function does not affect the location of the zeros. 

According to the well-established finite-size scaling theory, the shift of pseudocritical temperatures $T_{L}^{\ast}$ is described by the power law~\cite{ferrenberg91,amit_book}
\begin{equation}
T_{L}^{\ast}=T_{\rm c}+bL^{-1/\nu}(1+b'L^{-\omega}),
\label{eq:tc}
\end{equation}
where $T_{\rm c}$ is the critical temperature of the infinite system, $b$ and $b'$ are non-universal parameters, $\nu$ is the critical exponent of the correlation length, and $\omega$ denotes the correction-to-scaling (Wegner) exponent, fixed hereafter to the predicted value $\omega = 2$~\cite{alcaraz97,alcaraz99}. On the same ground, one 
also expects that~\cite{costa17}
\begin{equation}
x_{L}^{\ast}=x_{\rm c}+bL^{-1/\nu}(1+b'L^{-\omega}).
\label{eq:xc}
\end{equation}
Since $x_L^\ast \approx (1,0)$, the imaginary part $\operatorname{Im} (x_L^\ast)$ should 
scale with the system size as~\cite{costa17}
\begin{equation}
\operatorname{Im}(x_L^\ast) \sim L^{-1/\nu}(1+b'L^{-\omega}).
\label{eq:imx}
\end{equation}
In this description, the standard process is to firstly compute the critical exponent $\nu$ via Eq.~(\ref{eq:imx}), and then retrieve the critical temperature $T_{\rm c}$ using the Eq.~(\ref{eq:tc}).

\subsection{Results}
\label{sec:EPD_results}

\begin{figure}
\includegraphics[width=0.99\hsize]{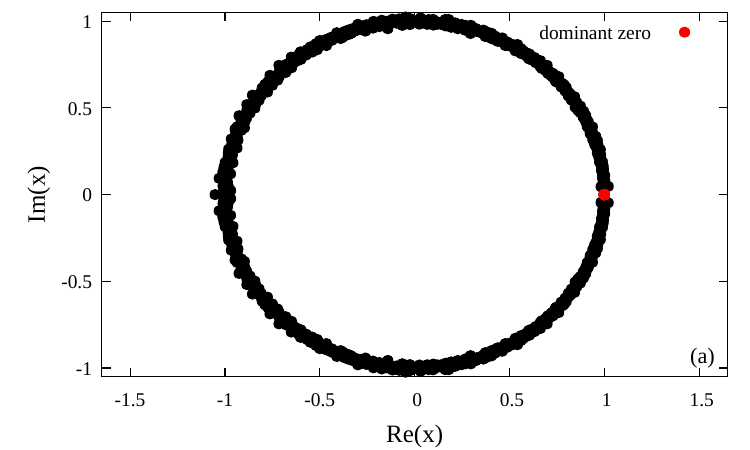}
\includegraphics[width=0.99\hsize]{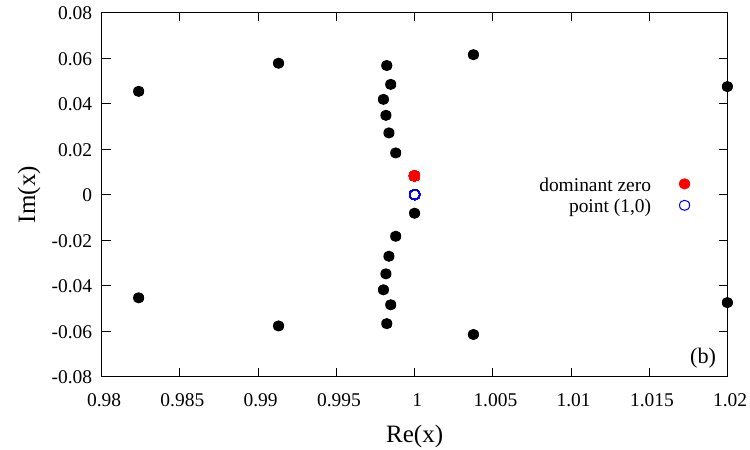}
\caption{Zeros of the EPD in the complex plane for the spin-$1/2$ Baxter-Wu model and 
a system with linear size $L = 75$ at $\beta = 0.43991$, close to the corresponding pseudocritical temperature; see also Eq.~(\ref{eq:hzeros}). Panel (a) illustrates a global view of the zeros around a unit-radius circle and panel (b) a zoom in on the area of the dominant zero near $(1,0)$. As discussed in the main text the roots appear in conjugate pairs.}
\label{fig:root}
\end{figure}
\begin{figure}
\includegraphics[width=0.99\hsize]{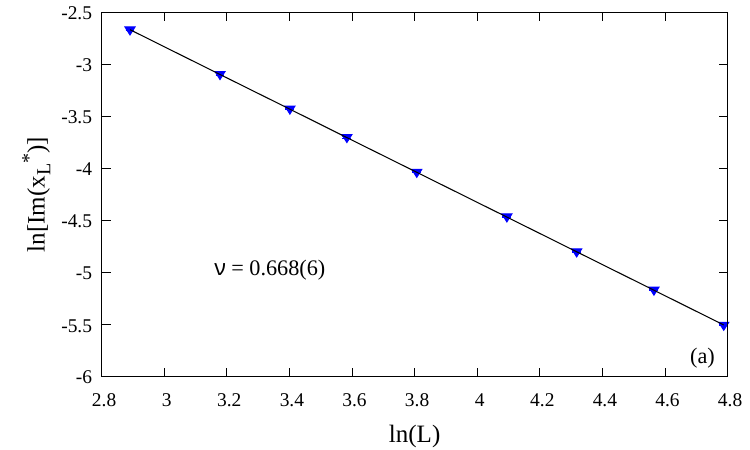}
\includegraphics[width=0.99\hsize]{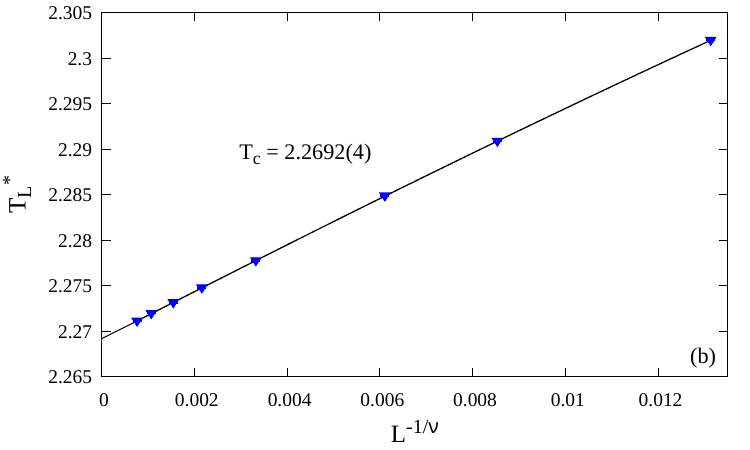}
\caption{(a) Log-log plot of the imaginary part of the dominant zero as a function of the lattice size for the spin-$1/2$ Baxter-Wu model. The solid line shows a fit of the form~(\ref{eq:imx}). (b) Finite-size scaling analysis of the pseudocritical temperatures, see Eq.~(\ref{eq:tc}), where the critical exponent $\nu$ is fixed to the value from panel (a).}
\label{fig:tcnus05}
\end{figure}

\begin{figure}
\includegraphics[width=0.99\hsize]{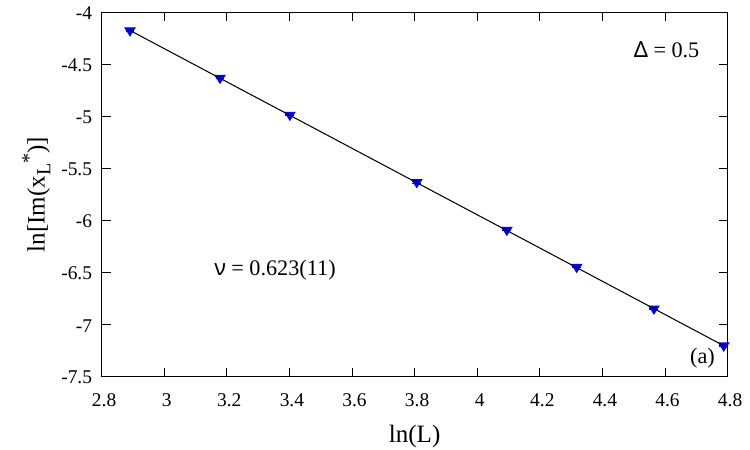}
\includegraphics[width=0.99\hsize]{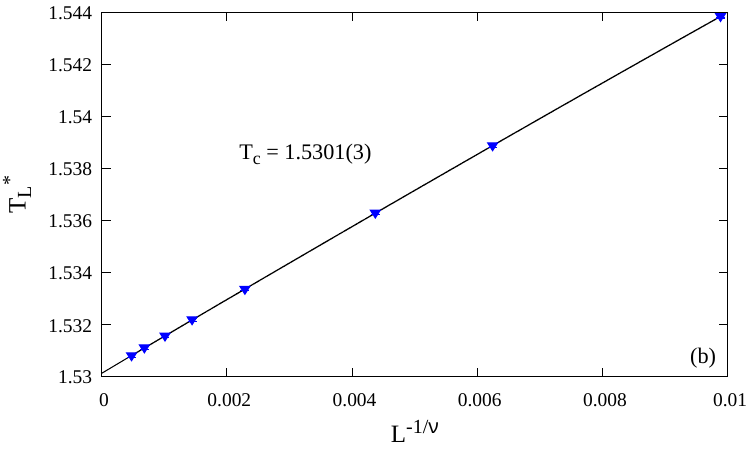}
\caption{The same as in Fig.~\ref{fig:tcnus05} for the spin-$1$ Baxter-Wu model at $\Delta=0.5$.}
\label{fig:d05}
\end{figure} 

For the application of the EPD zeros method to the Baxter-Wu model, histograms were accumulated using the standard single-spin-flip Metropolis algorithm. To accommodate for all ground states, periodic boundary conditions must be considered and the allowed values of the linear size of the lattice $L$ must be a multiple of three~\cite{dias17,costa04,fytas22,vasilopoulos22}. (Note that a triangular lattice on the torus is tripartite when its linear dimensions are multiples of
three). In the course of our simulations we considered linear sizes in the range $12 \le L \le 120$, respecting this constraint (a practise followed also in the multicanonical and hybrid simulations described below). During thermalization, $10^5$ Monte Carlo steps per spin (sweeps) were discarded for $L\le 45$ and $3\times 10^5$ sweeps for the larger sizes. An additional $10^8$ sweeps were performed to accumulate the energy histograms, leading to a quite precise estimate of the dominant root. 
The iterative process of finding the dominant EPD zero terminated when the temperature difference between two consecutive steps became smaller than a predefined accuracy of  $\varepsilon=10^{-4}$. Note that one may also look at the real part of the dominant zero and halt the process when $|\operatorname{Re}(x_L^\ast)-1| \le \varepsilon$, and also that smaller 
values of $\varepsilon$ may be considered,  without any significant consequences in the results. Regarding the cutoff, the value $h_{\rm cut}=10^{-4}$ was used throughout the simulations. Errors have been computed by averaging over $10$ different independent runs. Finally, for all fits performed throughout this paper we restricted ourselves to data with $L\geq L_{\rm min}$, adopting the standard $\chi^{2}$ test for goodness of the fit. Specifically, we considered a fit as being acceptable only if $10\% < Q < 90\%$, where $Q$ is the quality-of-fit parameter~\cite{press92}. 

It is clear that finding the zeros of a high-order polynomial is far from trivial, in general.  In the present case, this difficulty is being added to by the need to determine the cutoff of the EPD while monitoring the precision of coefficients $h_{\beta^j_0}(n)$ in order to obtain a sensible accuracy for the zeros. The precision of the coefficients $h_{\beta^j_0}(n)$ strongly depends on the length of the Monte Carlo time series. On the other hand, even in case of rather accurate values of the coefficients results still depend on the cutoff threshold of the EPD. However, as it has been recently shown for the two-dimensional Ising and $6$-state Potts models~\cite{rodrigues22}, the EPD method is indeed quite robust against the cutoff threshold and the number of Monte Carlo sweeps used, giving accurate results for the critical parameters.

Since the method has not yet been checked on the spin-$1/2$ Baxter-Wu model, our first port of call is to test it against the well-known exact results. Figure~\ref{fig:root} depicts typical results for the zeros of a system $L = 75$ at a temperature close to its pseudocritical. Note that since the density-of-states factors $g(E_n)$ are real, the zeros all come in conjugate pairs. The upper panel shows a global view of all zeros located around a unit circle, and the bottom panel depicts an enlargement of the dominant zero $(x^j)^\ast$ near the area $x_{\rm c} = (1,0)$. Changing the temperature according to Eq.~(\ref{eq:iter}) will furnish a different new dominant zero that converges to the desired $x_L^\ast$ after just a few iterations. The finite-size scaling analysis of the imaginary part of the dominant root, as given by Eq.~(\ref{eq:imx}), is shown in Fig.~\ref{fig:tcnus05}(a). A linear fit in a log-log scale gives an estimate of $\nu=0.668(6)$ for the critical exponent of the correlation length, in very good agreement with the exact result $\nu = 2/3$~\cite{baxter73,baxter_book}. Fixing $\nu$ to this value and $\omega = 2$~\cite{comment_omega}, Eq.~(\ref{eq:tc}) gives $T_{\rm c} = 2.2692(4)$ in accordance with the exact result $2.269185\cdots$~\cite{baxter_book}, see Fig.~\ref{fig:tcnus05}(b). 

\begin{figure}
\includegraphics[width=0.99\hsize]{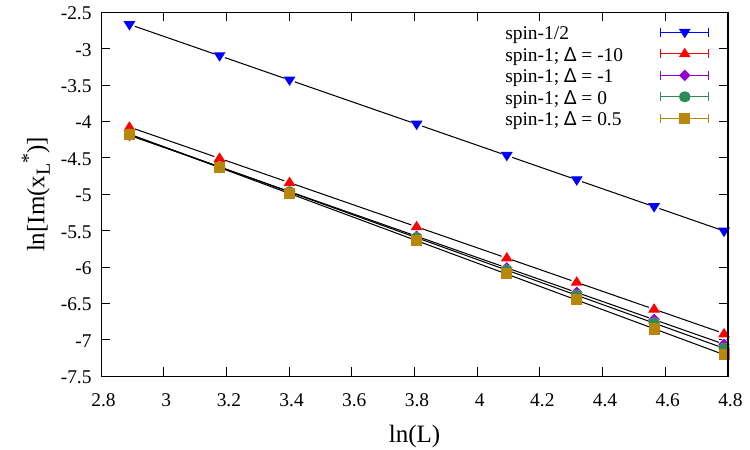}
\caption{Log-log plot of the imaginary 
part of the dominant zero as a function of the lattice size for both the spin-$1/2$ and spin-$1$ Baxter-Wu models. For the spin-$1$ case results at various values of $\Delta$ are shown. The solid lines are fits of the form~(\ref{eq:imx}) as described in the text.}
\label{fig:nud}
\end{figure}

We proceed now with the study of the spin-$1$ Baxter-Wu model at $\Delta=\{-10, -1, 0, 0.5\}$. This selection allows a direct comparison with results already reported in the literature by other approaches~\cite{dias17,jorge20,fytas22,vasilopoulos22}. For brevity, we chose to show here in Fig.~\ref{fig:d05} the case $\Delta = 0.5$, which is the largest positive value of $\Delta$ considered in this work. The scaling analysis in both panels of Fig.~\ref{fig:d05} is in direct analogy with that of Fig.~\ref{fig:tcnus05}, giving $\nu = 0.623(11)$ and $T_{\rm c} = 1.5301(3)$. Although the estimate for $T_{\rm c}$ is in excellent agreement with conformal invariance, see Tab.~\ref{tab:summary}, the value of $\nu$ appears to deviate from the expected $2/3$ result. A similar but slighter deviation was observed also for the case $\Delta = 0$, see again Tab.~\ref{tab:summary}. This trend is a note of warning indicating the presence of strong finite-size effects as $\Delta$ approaches the location $\Delta_{\rm pp}$ of the multicritical point, suggesting the need of studying larger system sizes. Finally, in Fig.~\ref{fig:nud} we provide a summary concerning the finite-size scaling behavior of the imaginary part of the dominant zero for all values of $\Delta$ considered, including the case of the spin-$1/2$ model. 
Inspecting Fig.~\ref{fig:nud} one may observe that as we lower $\Delta$ from $0.5$ to $-10$ the trend of the numerical data follows the expected passage to the spin-$1/2$ model ($\Delta = -\infty$). However, this approach appears to be rather slow, and it could instructive to study even more negative values of $\Delta$. The gathered results for $T_{\rm c}$ and $\nu$ are listed in Tab.~\ref{tab:summary} and are critically discussed in Sec.~\ref{sec:conclusions}. Overall, we may deduce that the EPD zeros method appears to be a promising alternative for determining critical aspects of the transition in the Baxter-Wu model.

\section{Multicanonical simulations}
\label{sec:muca}

\subsection{Method and observables}
\label{sec:muca_method}

\begin{figure}
\includegraphics[width=0.99\hsize]{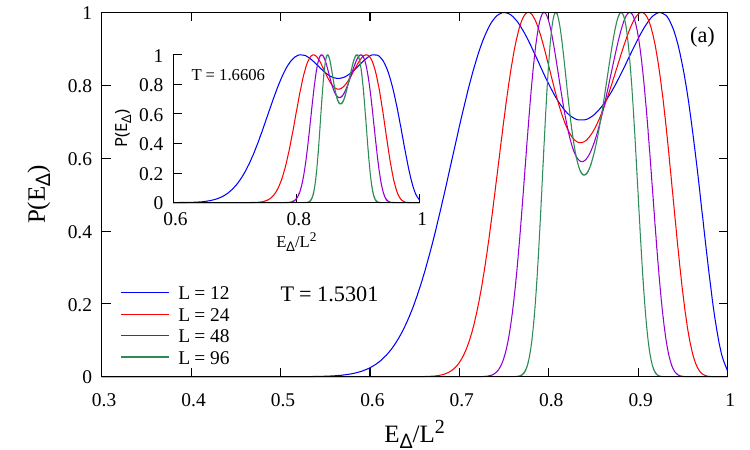}
\includegraphics[width=0.99\hsize]{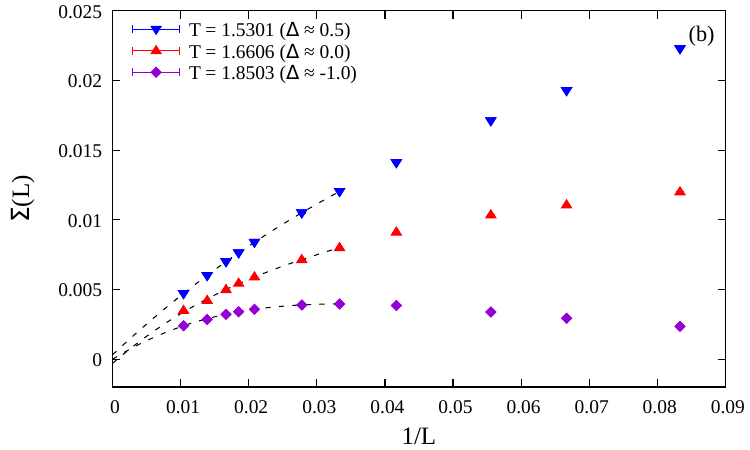}
\caption{(a) Reweighted canonical probability density functions $P(E_{\Delta})$ for selected system sizes at $T = 1.5301$ (main panel) and $T = 1.6606$ (inset). (b) Limiting behavior of the corresponding surface tension $\Sigma(L)$. For a better comparison of the pseudo-first-order effects, data at $T = 1.8503$ (corresponding approximately to $\Delta = -1$) are also included~\cite{vasilopoulos22}.}
\label{fig:muca_tension}
\end{figure}

\begin{figure}
\includegraphics[width=0.99\hsize]{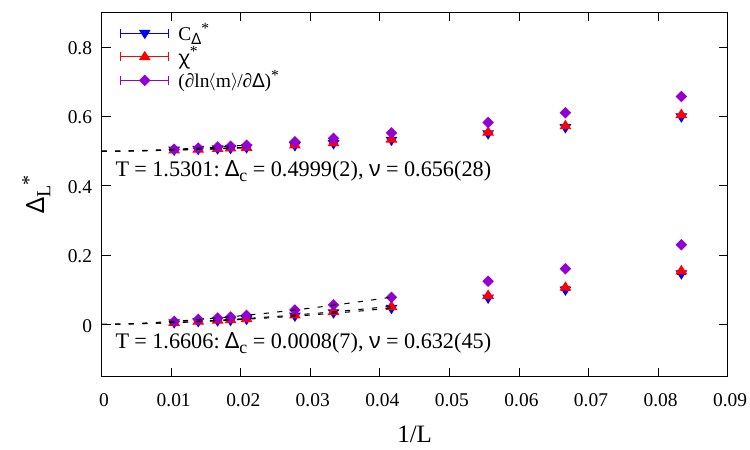}
\caption{Shift behavior of several pseudocritical fields as a function of the inverse linear system size at the temperatures $T = 1.5301$ and $T=1.6606$, corresponding roughly to $\Delta \approx 0.5$ and $\Delta \approx 0$, respectively.}
\label{fig:muca_D}
\end{figure}

The multicanonical (MUCA) method~\cite{berg92} consists of a substitution of the Boltzmann factor $e^{-\beta E}$ with weights that are iteratively modified to produce a flat histogram, usually in energy space. This ensures that suppressed states such as those in the co-existence region in an (effectively) first-order transition scenario can be reliably sampled, and a continuous reweighting to arbitrary values of the external control parameter becomes possible~\cite{janke03,gross18}. Due to the two-parametric nature of the density of states, $g(E_J, E_\Delta)$, in the spin-$1$ Baxter-Wu model, the process was applied only to the crystal-field part $E_\Delta$ of the energy. This allowed us to reweight to arbitrary values of $\Delta$ while keeping the temperature fixed. 
Starting from the partition function of Eq.~(\ref{eq:zn}) we can write
\begin{equation}
    {\cal Z}_\mathrm{MUCA}=\sum_{E_J, E_\Delta} g(E_J, E_\Delta)e^{-\beta E_J} W\left(E_\Delta\right),
\label{eq:zng}
\end{equation}
where the Boltzmann weight associated with the crystal-field part of the energy has been generalized to $W\left(E_\Delta\right)$. For a flat marginal distribution in $E_\Delta$, it should hold that
\begin{equation}
    W(E_{\Delta}) \propto {\cal Z}_\mathrm{MUCA} \left[ \sum_{E_J} g(E_J, E_\Delta) e^{-\beta E_J}  \right]^{-1}.
\label{eq:wmuca}
\end{equation}
In order to iteratively approximate the generalized weights $W\left(E_\Delta\right)$, we sampled histograms of the crystal-field energy. Supposing that at the $n^\text{th}$ iteration a histogram $H^{(n)}(E_\Delta)$ was sampled, then its average should depend on the weight of the iteration $W^{(n)}\left(E_\Delta\right)$ as
\begin{equation}
    \langle H^{(n)}(E_\Delta)\rangle \propto \sum_{E_J} g(E_J, E_{\Delta})e^{-\beta E_{J}} W^{(n)}(E_\Delta).
\label{eq:hmuca}
\end{equation}
From Eqs.~(\ref{eq:wmuca}) and (\ref{eq:hmuca}) it follows that $\langle H^{(n)}(E_\Delta)\rangle\propto W^{(n)}(E_\Delta) / W(E_\Delta)$. Hence, in order to approximate the $W(E_\Delta)$ that produces a flat histogram a weight modification scheme of the form $W^{(n+1)}\left(E_\Delta\right) = W^{(n)}\left(E_\Delta\right)/H^{(n)}(E_\Delta)$ is justified. The simulations can terminate when a flat-enough histogram has been sampled, based on a suitable flatness criterion. For our purposes we used the Kullback-Leibler divergence to test the flatness~\cite{kullback51, gross18}.
After this initial preparatory part, the final fixed weights can be used for production runs.

As has been shown in detail in Refs.~\cite{gross18,zierenberg13}, the multicanonical method can be adapted for the use on parallel machines by  performing the sampling of histograms in parallel, with each parallel worker using the same weights but a different (independent) pseudorandom number sequence. The accumulated histogram can then be used to update the weights, keeping communication between the parallel parts of the code minimal. This scheme has been successfully applied for the study of spin systems in the past, including the spin-$1$ Blume-Capel and Baxter-Wu models~\cite{fytas22,vasilopoulos22,zierenberg15,zierenberg17,fytas18}. Here we performed our simulations on an Nvidia Tesla K80 GPU, using a total of $26\; 624$ workers assigned to independent copies of the system. At each time, a subset of these threads are actually running in parallel on the $4\,992$ cores of the device, while the excess in the number of parallel tasks is employed to hide the latencies due to memory accesses~\cite{weigel18}. 

\begin{figure}
\includegraphics[width=0.99\hsize]{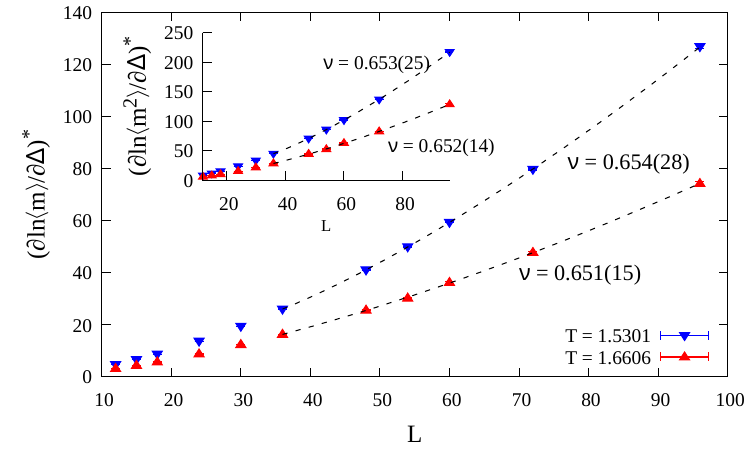}
\caption{Finite-size scaling behavior of the logarithmic derivatives (\ref{eq:log_der}) of powers $n=1$ (main panel) and $2$ (inset) of the order parameter.}
\label{fig:muca_dlnm}
\end{figure}

In the course of the multicanonical simulations 
the sampled observables include estimates of the mean energy $\langle E \rangle$, the order parameter $\langle m\rangle$ which is estimated from  the root mean square average of the magnetization per site of the three sublattices A, B, and C~\cite{jorge20,costa04b,costa16},
\begin{equation}\label{eq:order-parameter}
	m = \sqrt{\frac{m_{\rm A}^{2} +m_{\rm B}^{2} +m_{\rm C}^{2}}{3}},
\end{equation}
and the magnetic susceptibility 
\begin{equation}\label{eq:susceptibility}
 \chi = \beta N\left[\langle m^{2}\rangle - \langle m\rangle^{2}\right].
\end{equation}
As the multicanonical method allows for continuously reweighting to any value of $\Delta$, canonical expectation values for an observable $O =O (\{\sigma\})$ at a fixed temperature can be attained by estimating the expressions
\begin{equation}
	\langle O\rangle_\Delta
	=
	\frac{ \left\langle O(\{\sigma\})\,e^{-\beta\Delta E_{\Delta}(\{\sigma\})} W^{-1}\left(E_{\Delta}\right) \right\rangle_\mathrm{MUCA}}
	{ \left\langle e^{-\beta\Delta E_{\Delta}(\{\sigma\})} W^{-1}\left(E_{\Delta}\right) \right\rangle_\mathrm{MUCA} }.
	\label{eq:muca-reweight}
\end{equation}
In this framework, it is natural to compute $\Delta$-derivatives of observables rather than the usual $T$-ones. For instance, in place of the usual specific heat one may define a specific-heat-like quantity~\cite{zierenberg15}
\begin{equation}
	C_\Delta = \frac{1}{N} \frac{\partial E_J}{\partial\Delta}
	=
	-\beta\left[ \left\langle E_J E_\Delta \right\rangle - \left\langle E_J \right\rangle \left\langle E_\Delta \right\rangle \right]/N,
\end{equation}
which shows the shift behavior expected from the usual specific heat~\cite{fytas22,vasilopoulos22,zierenberg15}. Additionally, in order to obtain direct estimates of the critical exponent $\nu$ from finite-size scaling, one may compute the logarithmic derivatives of $n$th power of the order parameter~\cite{ferrenberg91,caparica00,malakis09}
\begin{equation}
\label{eq:log_der}
	\frac{\partial\ln{\langle m^n \rangle}}{\partial \Delta}
	=
	-\beta\left[\frac{\left\langle m^n E_\Delta \right\rangle}{\langle m^n \rangle} - \left\langle E_\Delta \right\rangle  \right].
\end{equation}

\begin{figure}
\includegraphics[width=0.99\hsize]{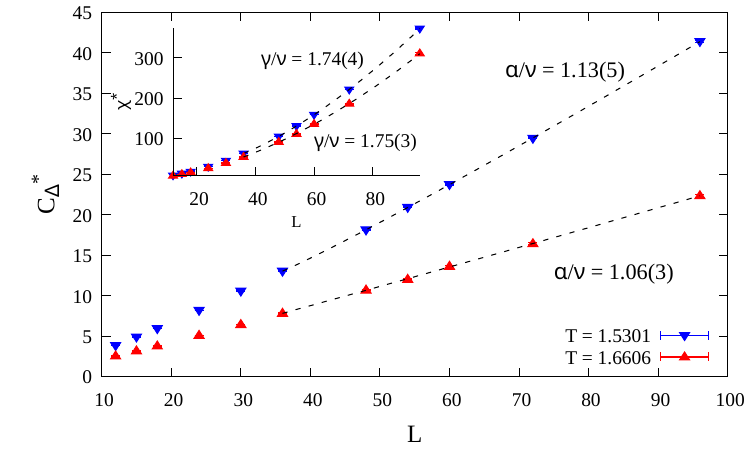}
\caption{Finite-size scaling behavior of $C_{\Delta}^{\ast}$ (main panel) and $\chi^{\ast}$ (inset).}
\label{fig:muca_cx}
\end{figure}

\begin{table*}[]
\caption{Representative critical-point estimates ($T_{\rm c}$ or $\Delta_{\rm c}$) of the phase diagram of the spin-$1$ Baxter-Wu model from the present work and as well as previous studies, including estimates of the critical exponent $\nu$. In the first column we either indicate the value of $\Delta$ for simulations that vary $T$ or the value of $T$ for simulations that vary $\Delta$. Columns 2 -- 5 feature results obtained in the current work from the EPD zeros method (columns 2 and 3) and multicanonical simulations (columns 4 and 5). Columns 6 -- 9 append earlier estimates from Wang-Landau (WL) simulations ($\Delta = -10$ and $-1$~\cite{vasilopoulos22} and $\Delta = 0$~\cite{jorge20}) and conformal invariance (CI)~\cite{dias17}. The first line of results relates to the pure spin-$1/2$ model ($\Delta = -\infty $).}
\vspace*{0.2cm}
\centering
\label{tab:summary}
\begin{tabular}{l|ccc|ccc|ccc|ccc}
\hline\hline
                                   &              & EPD Zeros  &           &                  & MUCA &            &             & WL &             &             & CI &         \\
Simulation  point                & $T_{\rm c}$  &      & $\nu$   & $\Delta_{\rm c}$ &      & $\nu$      & $T_{\rm c}$ &    & $\nu$       & $T_{\rm c}$ &    & $\nu$  \\ \hline\hline
~~~$\Delta=-\infty$   & $2.2692(4)$ &  & $0.668(6)$ & --               &      & --         & --          &    & --          & --          &    & --      \\ \hline
~~~$\Delta=-10$                    & $2.2578(4)$ &  & $0.666(10)$ & --               &      & --         & $2.2578(5)$ &    & $0.655(17)$ & $2.2578$    &    & $0.6683$ \\ \hline
~~~$\Delta=-1$                     & $1.8502(3)$ &  & $0.660(12)$ & --               &      & --         & $1.8503(9)$ &    & $0.652(18)$ & $1.8503$    &    & $0.6601$ \\
~~~$T=1.8503$                    & --           &          &    --       & $-1.002(2)$      &      & $0.671(6)$\footnote{From Ref.~\cite{vasilopoulos22}.} & --          &    & --          & --          &    & --  \\ \hline
~~~$\Delta=0$                      & $1.6606(5)$ &  & $0.649(12)$ & --               &      & --         & $1.66055(5)$          &    & $0.644(1)$          & $1.6606$    &    & $0.6488$ \\
~~~$T=1.6606$                     & --           &         &       --     & $0.0008(7)$      &      & $0.652(10)$\footnote{This estimate (and the one at $T = 1.5301$) corresponds to the average value of $\nu$ obtained from the fits of Fig.~\ref{fig:muca_dlnm}. Cross-correlations were not taken into account, but see Ref.~\cite{weigel09}.}  & --          &    & --          & --          &    & --  \\ \hline
~~~$\Delta=0.5$                    & $1.5301(3)$ &  & $0.623(11)$ & --               &      & --         & --          &    & --          & $1.5300$    &    & $0.6369$\footnote{Private communication by the authors of Ref.~\cite{dias17}.} \\
~~~$T=1.5301$                     & --           &         &     --       & $0.4999(2)$      &      & $0.654(19)$ & --           &    & --          & --          &    & -- \\
\hline\hline
\end{tabular}
\end{table*}

\subsection{Results}
\label{sec:muca_results}

We performed simulations at $T = 1.6606$ and $T = 1.5301$ which approximate the critical points at $\Delta = 0$ and $\Delta = 0.5$ respectively, see Table~\ref{tab:summary}, using system sizes $12 \le L \le 96$, again with periodic boundary conditions. For $T = 1.6606$, $4\times 10^6$ sweeps were used in the production run for the smallest system and $3\times 10^8$ sweeps for the largest. For the lower temperature $T = 1.5301$, due to its proximity to the proposed multicritical point (see Fig.~\ref{fig:phase_diagram}), sampling was increased to $2.5\times 10^7$ sweeps in the production run for the smallest system and $10^9$ sweeps for the largest. After the initial iterations for the calculation of the generalized weights, an additional $10\%$ of the total production sweeps were discarded by each worker for thermalization. Preliminary tests indicated that distributing the production sweeps equally among the workers results in sampling the equivalent of $\sim 5$ autocorrelation times worth of data points per worker for the larger temperature and $\sim 20$ for the smaller. The results were analyzed using the jackknife resampling method~\cite{efron} and the location of pseudocritical points was estimated via reweighting and bisecting in $\Delta$.

As discussed in Refs.~\cite{vasilopoulos22,jorge21,jorge20}, there have been recent reports of first-order transition features even along the presumed continuous part of the transition line. In relation to such claims, we put forward here some additional evidence for the clarification of the nature of the phase transition at $\Delta < \Delta_{\rm pp}$. Following the prescription of Ref.~\cite{vasilopoulos22} we studied the reweighted probability density function $P(E_{\Delta})$. It is well known that a double-peak structure in the density function in finite systems is an expected precursor of the two $\delta$-peak behavior in the thermodynamic limit that is expected for a first-order phase transition~\cite{binder84,binder87}. However, this observation must be taken with a grain of salt, since there have been many cases reported in the literature, for which this two-peak structure tends to a unique peak in the thermodynamic limit. A warning example is the two-dimensional $4$-state Potts model~\cite{fernandez09}.

We start the presentation of our results with Fig.~\ref{fig:muca_tension}(a) 
where we show the probability density function $P(E_{\Delta})$ for selected system sizes at the temperatures $T = 1.6606$ and $1.5301$. A double-peak structure is observed in both cases, in agreement with the evidence in Ref.~\cite{jorge20} for $\Delta = 0$. As is clearly visible, stronger first-order-like characteristics are present for the lower-$T$ (higher-$\Delta$) example that is closer to the multicritical point.

The multicanonical method is optimal for studying these phenomena in the framework of the method proposed by Lee and Kosterlitz~\cite{lee90}, as it allows the direct estimation of the barrier associated with the suppression of states during a first-order phase transition. Considering distributions with two peaks of equal height (eqh)~\cite{borgs92}, as the ones shown in Fig.~\ref{fig:muca_tension}(a), allows one to extract the surface tension in the $E_{\Delta}$-space,
\begin{equation}
\label{eq:surface_tension}
	\Sigma (L)  = \frac{1}{2\beta L}\ln\left(\frac{P_{\rm
			max}}{P_{\rm min}}\right)_{\rm eqh},
\end{equation}
where $P_{\rm max}$ and $P_{\rm min}$ are the maximum and
local minimum of the distribution $P(E_{\Delta})$, respectively. This parameter is expected to scale in two dimensions as 
\begin{equation}
\label{eq:surface_tension_scaling}
\Sigma(L)  = \Sigma_{\infty} + c_1L^{-1} + c_2L^{-2} + c_3L^{-3},
\end{equation}
possibly with higher-order corrections~\cite{Nussbaumer2006,Nussbaumer2008,Bittner2009}. For the system under investigation here, the scaling behavior of the surface tension $\Sigma(L)$ is depicted in Fig.~\ref{fig:muca_tension}(b) for all temperatures studied. The dashed lines show fits of the form~(\ref{eq:surface_tension_scaling}) with $L\geq L_{\rm min} = 30$ leading to a practically zero value of $\Sigma_{\infty}$ in all cases. In particular, we obtain the extrapolated values $\Sigma_{\infty} = -0.00005(11)$, $-0.0003(9)$, and $0.0003(3)$, for $T = 1.8503$, $1.6606$, and $1.5301$, respectively. This analysis suggests a continuous transition in the thermodynamic limit for the regime of $\Delta < \Delta_{\rm pp}$, in favor of the scenario originally discussed in Ref.~\cite{vasilopoulos22}.

In order to extract critical crystal fields $\Delta_{\rm c}(T)$ as well as a first estimate of the correlation-length exponent $\nu$, we present in Fig.~\ref{fig:muca_D} the shift behavior of suitable pseudocritical fields $\Delta_{L}^{\ast}$. These are defined as the peak locations of $\Delta$-dependent curves, such as the specific heat $C_{\Delta}$, the magnetic susceptibility $\chi$, and the logarithmic derivative of the order parameter $\partial\ln{\langle m \rangle}/\partial \Delta$. For each of the two temperatures studied the dashed lines show joint fits to the expected power-law behavior~\cite{zierenberg15,zierenberg17} 
\begin{equation}\label{eq:pseudo_Delta_scaling}
	\Delta^{\ast}_{L}=\Delta_{\rm c}+bL^{-1/\nu}(1+b'L^{-\omega}),
\end{equation}
where $\Delta_{\rm c}$ and $\nu$ are common parameters and $\omega = 2$~\cite{alcaraz97,alcaraz99}. Using $L_{\rm min} = 24$ and $48$ for $T = 1.6606$ and $1.5301$, respectively, the evaluated critical points $\Delta_{\rm c} (T = 1.5301) = 0.4999(2)$ and $\Delta_{\rm c} (T = 1.6606)= 0.0008(7)$ are in good agreement with the results of Sec.~\ref{sec:EPD_results} but also with those reported in Tab.~\ref{tab:summary} from Wang-Landau simulations~\cite{jorge20} and conformal invariance~\cite{dias17}. More importantly, our estimates $\nu = 0.656(28)$ and $0.632(45)$ for $T = 1.5301$ and $1.6606$, respectively, confirm to a good accuracy the $q=4$ Potts model universality class~\cite{wu82}.  

Additional estimates for the critical exponent $\nu$ can be extracted from the maxima of the logarithmic derivatives of the order parameter according to Eq.~(\ref{eq:log_der}), which are expected to scale as~\cite{ferrenberg91,caparica00,malakis09}
\begin{equation}\label{eq:log_der_scaling}
	\left(\frac{\partial\ln{\langle m^n \rangle}}{\partial \Delta}\right)^{\ast} \sim L^{1/\nu}(1+b'L^{-\omega}).
\end{equation}
Figure~\ref{fig:muca_dlnm} shows our data for $n=1$ (main panel) and $n=2$ (inset) at the two temperatures under study. The dashed lines are power-law fits of the form~(\ref{eq:log_der_scaling}) using $L_{\rm min} = 36$, providing an average of $\nu = 0.654(19)$ and $0.652(10)$ for $T = 1.5301$ and $1.6606$, respectively, thus reinforcing the scenario of the $q=4$ Potts model universality class~\cite{domany78}. 

\begin{figure}[tb!]
\includegraphics[width=0.99\hsize]{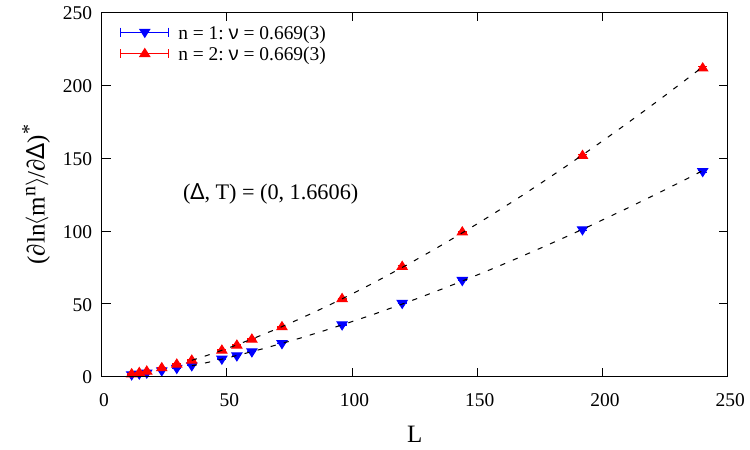}
\caption{Finite-size scaling of the logarithmic derivatives of powers $n = 1$ and $2$ of the order parameter at the critical point $(\Delta, T) = (0, 1.6606)$, as indicated by the EPD zeros method; see also Tab.~\ref{tab:summary}. The dashed lines are fits of the form~(\ref{eq:log_der_scaling}). Data generated via the hybrid approach.}
\label{fig:hybrid_dlnm}
\end{figure}

\begin{table}[tb!]
\caption{Summary of results for the critical exponent $\nu$ of the spin-$1$ Baxter-Wu model at $\Delta = 0$ obtained via conformal invariance (2\textsuperscript{nd} row)~\cite{dias17}, Wang-Landau simulations (3\textsuperscript{rd} row)~\cite{jorge20}, EPD zeros (4\textsuperscript{th} row), and multicanonical simulations (5\textsuperscript{th} row). The 6\textsuperscript{th} row showcases the estimate of $\nu$ via the hybrid approach~\cite{novotny93,vasilopoulos23}, see also Fig.~\ref{fig:hybrid_dlnm}. For all methods the maximum accessible system size $L_{\rm max}$ used in the simulations is also given in brackets. The 3\textsuperscript{rd} column highlights the deviation $\delta \nu$ of each estimate from the exact value~\cite{wu82}, which is included in the last row for reference.}
\vspace*{0.2cm}
\label{tab:D0}
\begin{tabular}{|c|c|c|}
\hline 
Method & $\nu$ & $\delta \nu = |2/3 - \nu|$\\ \hline
\begin{tabular}[c]{@{}c@{}}CI\\ ($L_{\rm max} = 12$)\footnote{Note that $L_{\rm max}$ denotes the maximum strip width considered in Ref.~\cite{dias17}. }\end{tabular}     & $0.6488$  & $0.018$\\ \hline
\begin{tabular}[c]{@{}c@{}}WL\\ ($L_{\rm max} = 92$)\end{tabular}    & $0.644(1)$ & $0.023(1)$  \\ \hline
 \begin{tabular}[c]{@{}c@{}} EPD zeros\\ ($L_{\rm max} = 120$)\end{tabular}         & $0.649(12)$  & $0.018(12)$ \\ \hline
\begin{tabular}[c]{@{}c@{}}MUCA\\ ($L_{\rm max} = 96$)\end{tabular}                &   $0.652(10)$\footnote{Average value of $\nu$ obtained from the fits of Fig.~\ref{fig:muca_dlnm}.} &  $0.015(10)$ \\ \hline
\begin{tabular}[c]{@{}c@{}}Hybrid\\ ($L_{\rm max} = 240$)\end{tabular}              & $0.669(3)$    &  $0.002(3)$\\ \hline
Exact solution  & $2/3$  & 0\\
 \hline
\end{tabular}
\end{table}

Finally, we turn to the finite-size scaling behavior of the maxima of the specific heat, $C^{\ast}_{\Delta}$, and magnetic susceptibility, $\chi^{\ast}$, in order to probe the critical-exponent ratios $\alpha/\nu$ and $\gamma / \nu$, respectively. Figure~\ref{fig:muca_cx} contains the relevant numerical data at the two temperatures considered. The dashed lines are fits of the expected form~\cite{zierenberg17,vasilopoulos22}
\begin{equation}\label{eq:spec-heat_scaling}
	C^{\ast}_{\Delta} \sim L^{\alpha/\nu}(1+b'L^{-\omega})
\end{equation}
and
\begin{equation}\label{eq:susceptibility_scaling}
	\chi^{\ast} \sim L^{\gamma/\nu}(1+b'L^{-\omega}),
\end{equation}
with $L_{\rm min} = 36$. These led to the estimates $\alpha/\nu = 1.13(5)$ and $1.06(3)$, and $\gamma/\nu = 1.74(4)$ and $1.75(3)$ for $T = 1.5301$ and $1.6606$, respectively. Here, at the lower temperature $T = 1.5301$ we had to include a second-order correction term ($\sim L^{-2\omega}$) in our fitting attempts to improve the quality-of-fit.
All of the above results are clearly compatible with the exact values $\alpha/\nu = 1$ and $\gamma/\nu = 7/4$ of the $4$-state Potts model universality class~\cite{wu82}. 

\section{Summary and outlook}
\label{sec:conclusions}

In closing, we return to the question of the current understanding of the behavior of the model along the phase boundary. Our results as well as some reference estimates from the recent literature are summarized in Table~\ref{tab:summary}. On inspecting these values, the following comments are in order: (i) A very good agreement between different methods of estimating the location of points ($\Delta$, $T$) along the phase boundary of the model is observed, cross-validating the different numerical approaches used in the present but also in previous works. (ii) The values of the critical exponent $\nu$ at $\Delta < 0$ are fully compatible with the value $2/3$ of the $4$-state Potts universality class~\cite{domany78,wu82} for all methods. However, with increasing $\Delta$, a slight decrease in the value of $\nu$ is observed and may be attributed to the presence of finite-size effects that become more pronounced as one approaches the pentacritical point $\Delta_{\rm pp} \approx 0.89 - 1.68$~\cite{dias17,jorge21}. (iii) Although the multicanonical simulations allowed us to significantly improve the limited capability of the Metropolis algorithm to reduce correlations, much larger system sizes are required for a safe determination of critical exponents, in particular in the regime $0 \leq \Delta < \Delta_{\rm pp}$. 

In light of the above discussion, it would be very valuable to have at one's disposal some simulation method that allows to equilibrate significantly larger systems than those considered here. This holds especially for the scaling at the pentacritical point itself, where one may need to take into account possible multiplicative and additive logarithmic corrections, similar to those present in the $4$-state Potts model. A suitable cluster update for the spin-$1/2$ Baxter-Wu model was proposed by Novotny and Evertz~\cite{novotny93}. Its basic idea is as follows: for each update step one of the sublattices is chosen at random and its spins kept fixed, resulting in an effective Ising model on the other two sublattices with non-frustrating couplings. Hence, the Swendsen-Wang~\cite{swendsen87} algorithm can be applied for simulations of these embedded models. Our preliminary tests indicate that a combination of this cluster formalism that improves the decorrelation of configurations but is not ergodic as it does not affects the diluted spins $\sigma_i = 0$ with the heat bath algorithm~\cite{miyatake86,loison04} results in an efficient algorihtm capable of thermalizing rather large systems. A detailed analysis of the critical dynamical behavior of this hybrid scheme will be presented elsewhere~\cite{vasilopoulos23}.

Here, we confine ourselves to an exemplary application of this technique to the case $\Delta = 0$ beyond which the deviation in the estimates of $\nu$ from the expected value $2/3$ appears to grow, cf.\ also Table~\ref{tab:summary}. In Fig.~\ref{fig:hybrid_dlnm} we present the results of a test calculation of the critical exponent $\nu$ from hybrid simulations at the critical point $(\Delta,T) = (0,1.6606)$, studying systems up to linear size $L_{\rm max} = 240$. The finite-size scaling analysis of the logarithmic derivatives of the order parameter (for both $n = 1$ and $2$) produces the estimate $\nu = 0.669(3)$, in excellent agreement with the value $2/3$~\cite{baxter73,domany78,wu82}. A comparative set of results for the critical exponent $\nu$ of the spin-$1$ Baxter-Wu model at $\Delta = 0$ is given in Tab.~\ref{tab:D0}, where one may notice the superior accuracy of the hybrid approach.

To conclude, this work complements previous results that map the universality class of the spin-$1$ Baxter-Wu model to that of the $4$-state Potts model, a nontrivial task obscured by the presence of strong finite-size effects as revealed by our analysis. Clearly, it would be very instructive to add data for additional values of  $\Delta$ in the regime $\Delta > 0.5$. Yet this would require a huge computational effort given that crossover phenomena become more pronounced as we move towards the expected pentacritical point. This indicates that in order to perform a safe finite-size scaling analysis much larger system sizes would be needed with increasing values of $\Delta$. For future work, we propose the following two-stage process: (i) identify with good numerical accuracy the location of the pentacritical point ($\Delta_{\rm pp}$, $T_{\rm pp}$) and (ii) perform extensive simulations around this point using the hybrid approach in order to quantify all these interesting phenomena outlined above, including 
crossover effects and possible logarithmic corrections to scaling. A possible tool for such an endeavor could be the field-mixing technique~\cite{wilding92} in combination with the numerical methods reported in this paper. Such attempts are the subject of ongoing investigations.

\begin{acknowledgments}
We would like to thank Prof. Lucas M\'ol for fruitful discussions on the use of the EPD zeros method and Prof. Gerald Weber for the invaluable assistance in the use of the Statistical Mechanics Computer Lab facilities at the Universidade Federal de Minas Gerais. We acknowledge the provision of computing time on the parallel computer clusters \emph{ZEUS} and \emph{EPYC} of Coventry University. This research was supported by CNPq, CAPES, and FAPEMIG (Brazilian agencies).
\end{acknowledgments}


\begin{thebibliography}{99}
\bibitem{comment_omega} Here, and in all fits shown below, we have used the exact value for the corrections-to-scaling exponent $\omega$, since our tests indicated that even if it is treated as an additional free fitting parameter in Eq.~(\ref{eq:tc}) it always converges to $\sim 2$. For these tests we have used again the standard $\chi^{2}$ test for goodness of the fit, as described in the main text.
\end{thebibliography}
\end{document}